\documentclass[aps,prl,twocolumn,showpacs,amsmath,amssymb,longbibliography,superscriptaddress]{revtex4-1} 
\usepackage{graphicx} 
\usepackage{dcolumn} 
\usepackage{bm} 

\begin{document}

\title{Propagating irreversibility fronts in cyclically-sheared suspensions}

\author{Jikai Wang}
\affiliation{Department of Physics, Syracuse University, Syracuse, NY 13244, USA}
\affiliation{BioInspired Syracuse: Institute for Material and Living Systems, Syracuse University, Syracuse, NY 13244, USA}
\author{J. M. Schwarz}
\email{jmschw02@syr.edu}
\affiliation{Department of Physics, Syracuse University, Syracuse, NY 13244, USA}
\affiliation{BioInspired Syracuse: Institute for Material and Living Systems, Syracuse University, Syracuse, NY 13244, USA}
\affiliation{Indian Creek Farm, Ithaca, NY 14850, USA}
\author{Joseph D. Paulsen}
\email{jdpaulse@syr.edu}
\affiliation{Department of Physics, Syracuse University, Syracuse, NY 13244, USA}
\affiliation{BioInspired Syracuse: Institute for Material and Living Systems, Syracuse University, Syracuse, NY 13244, USA}
\date{\today}

\begin{abstract}
The interface separating a liquid from its vapor phase is diffuse: the composition varies continuously from one phase to the other over a finite length. 
Recent experiments on dynamic jamming fronts in two dimensions [Waitukaitis \textit{et al.}, Europhysics Letters \textbf{102}, 44001 (2013)] identified a diffuse interface between jammed and unjammed discs. 
In both cases, the thickness of the interface diverges as a critical transition is approached. 
We investigate the generality of this behavior using a third system: a model of cyclically-sheared non-Brownian suspensions. 
As we sediment the particles towards a boundary, we observe a diffuse traveling front that marks the interface between irreversible and reversible phases. 
We argue that the front width is linked to a diverging correlation lengthscale in the bulk, which we probe by studying avalanches near criticality. 
Our results show how diffuse interfaces may arise generally when an incompressible phase is brought to a critical point. 
\end{abstract}

\maketitle

Whereas Young and Laplace conceived of fluid interfaces as having zero thickness, it is now understood that physical properties vary smoothly through them \cite{Anderson98}. 
This situation becomes most apparent near a critical point, where 
interfacial thicknesses diverge \cite{Cahn58,Huang69}. 
Recently, cyclically-sheared non-Brownian suspensions have emerged as a testbed for studying \textit{non-equilibrium} phase transitions~\cite{Pine05,Corte08,Menon09,Tjhung15}. 
This system exhibits a dynamically-reversible phase where particle trajectories retrace themselves in each cycle, and an irreversible phase where particle collisions lead to diffusive behavior~\cite{Pine05,Corte08,Xu13,Milz13,Schrenk15,Pham16}. 
It is natural to ask whether an interface between these phases may be produced, and if so, what its properties are. 
Moreover, because cyclic shear is emerging as a strategy for controlling rheological properties~\cite{Paulsen14,Lin16,Ness18}, this understanding could impact the industrial processing of suspensions, where particle concentration or shear strain often varies spatially, as in pipe flow \cite{Guasto10,Snook16}. 

\begin{centering}
\begin{figure}[b]
\includegraphics[width=0.88\linewidth]{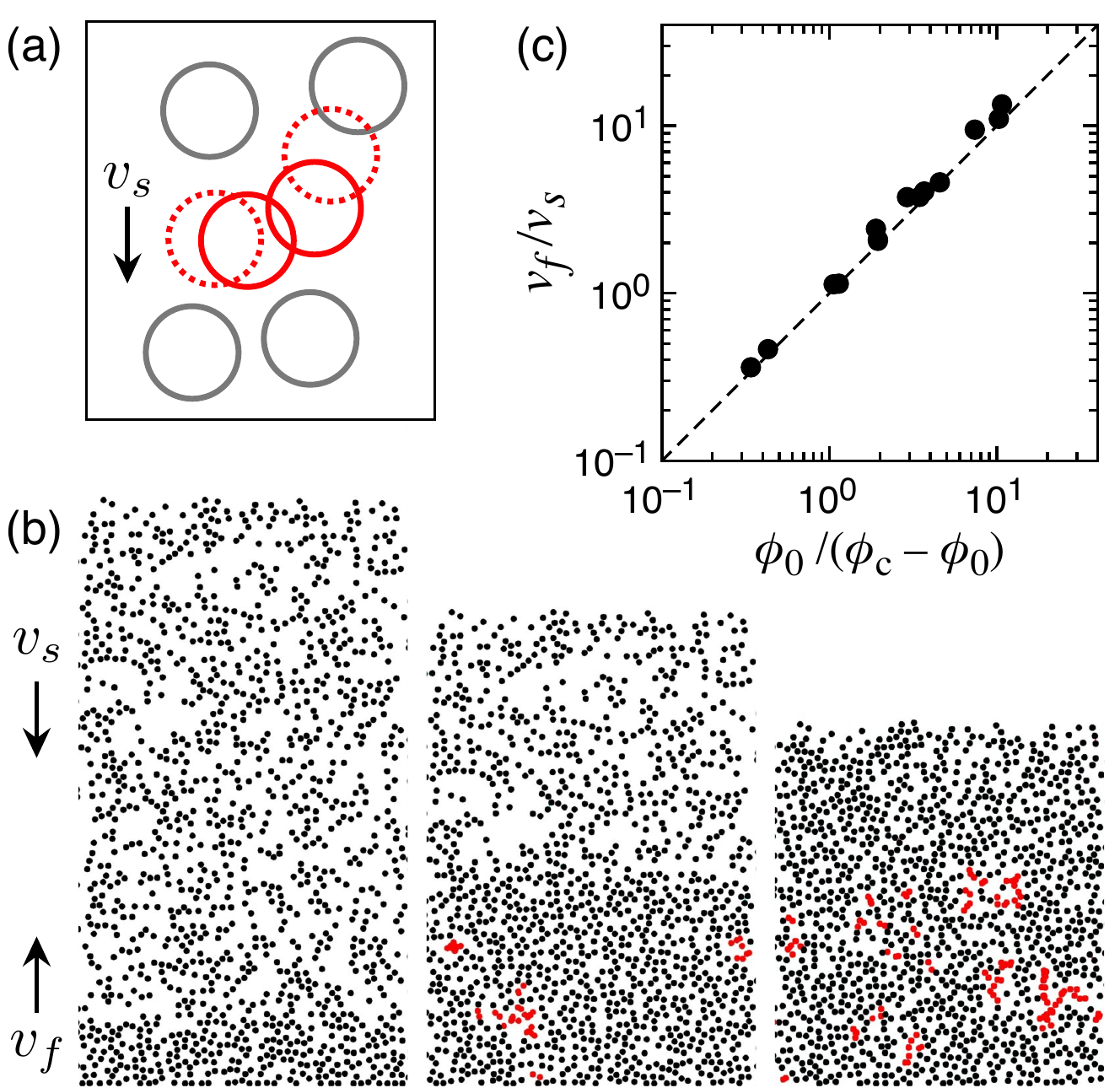}
\caption{
\textbf{Self-organized compaction front.} 
\textbf{(a)} Simplified model of a cyclically-sheared, sedimenting suspension after Ref.~\cite{Corte09}. 
In each cycle, a uniform sedimentation velocity $v_s$ is applied to all particles, and particles that overlap (red) are given random kicks. 
\textbf{(b)} Typical simulation showing a traveling front between a dense fluctuating region and a dilute reversible region. 
The front moves at constant speed $v_f$ until it reaches the top of the sediment and a fluctuating steady state begins. 
Here, $N=1273$, $\phi_0=0.2$, $\epsilon=0.5$, $W=50$, $H=100$, $v_s=2\times 10^{-5}$. 
\textbf{(c)} Scaled front velocity, $v_f/v_s$. 
The data over a wide range of parameters are well-described by Eq.~(\ref{eq:v_f}), which assumes the two phases have uniform densities equal to $\phi_0$ and $\phi_c$. 
Here, $0.05 \leq \epsilon \leq 10$; $300 < N < 16300$; $10^{-6} \leq v_s \leq 4 \times 10^{-4}$; $0.05 \leq \phi_0 \leq 0.40$; $0.16 < \phi_c < 0.46$. 
}
\label{fig:1}
\end{figure}
\end{centering}

Here we study the random organization of particles that are driven towards a hard boundary, using a simplified model of cyclically-sheared suspensions \cite{Corte09,Wang18}. 
This setup produces a well-defined interface between two bulk phases: a dense irreversible phase that builds up from the bottom wall, and a reversible sinking phase (Fig.~\ref{fig:1}b). 
We find that the interface has a finite thickness that diverges as the sinking phase approaches the critical density. 
We then link the interface thickness to a bulk correlation length by measuring a growing correlation length in systems without a sedimentation. 
Our results show strong similarities with dynamic jamming fronts~\cite{Waitukaitis12,Burton13}, where an interface between two non-equilibrium phases was identified with similar properties~\cite{Waitukaitis13}. 

\begin{centering}
\begin{figure*}[tb]
\includegraphics[width=0.92\linewidth]{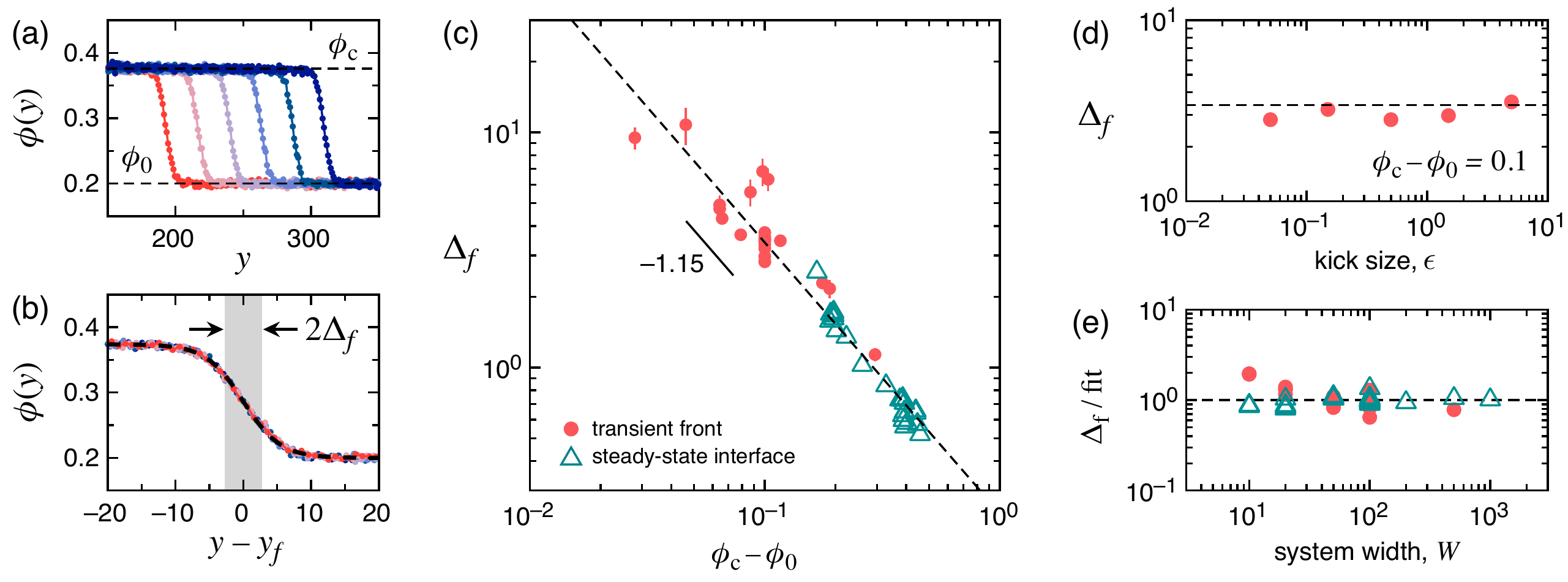}
\caption{
\textbf{Interface shape and thickness.} 
\textbf{(a)} Density profile snapshots for $\epsilon=0.5$, $N=16297$, $v_s=1.7\times10^{-5}$, $W=100$, sampled at a regular period. Each curve is averaged over $200$ systems. 
The data plateau to the dashed lines at $\phi_0=0.2$, $\phi_c=0.376$. 
\textbf{(b)} Translating these 6 profiles atop one another shows that the front moves with fixed shape and width. 
The profile is consistent with a sigmoid [dashed line: Eq.~(\ref{eq:phi_y})]. 
\textbf{(c)} Measured front width, $\Delta_f$, versus proximity to criticality of the sedimenting phase, $\phi_c - \phi_0$. 
Closed symbols: transient fronts. 
Open symbols: interface at the top of the system in the steady state (where $\phi_0 = 0$ above the sediment). 
The data are consistent with a power law with exponent $-1.15 \pm 0.18$ (dashed line), over a wide range of parameters ($0.05 \leq \epsilon \leq 10$; $300 < N < 16300$; $10^{-7} \leq v_s \leq 4 \times 10^{-4}$; $0.05 \leq \phi_0 \leq 0.40$; $0.16 < \phi_c < 0.46$). 
\textbf{(d)} The front width does not depend on $\epsilon$. 
Here we adjust $\phi_0$ so that $\phi_c - \phi_0 = 0.1$ is constant; all other parameters are fixed ($N=1730$, $v_s=10^{-6}$, $W=50$). 
Dashed line: value of the fit in panel (c). 
\textbf{(e)} Scaling the front width by the power law fit from panel (c), which shows that $\Delta_f$ does not depend strongly on the system width, $W$. 
}
\label{fig:2}
\end{figure*}
\end{centering}

\textit{Model---} Our simulations are based on a simplified model of cyclically-sheared suspensions proposed by Cort\'e \textit{et al.}~\cite{Corte08}, which evolves the positions of $N$ discs of diameter $d=1$ in a box of width $W$ and height $H$ using discrete cycles. 
We use an isotropic version of the model \cite{Tjhung15}, where particles that overlap in a cycle are given a small kick in a random direction (Fig.~\ref{fig:1}a), to emulate local irreversibility due to collisions \cite{Popova07}. 
The kick magnitude is chosen uniformly between $0$ and $\epsilon$, which we vary from $0.05$ to $10$. 
For small area fractions $\phi_0 = N\pi/(4WH)$, 
the system self-organizes into one of many absorbing states, where there are no overlaps and the dynamics is reversible thereafter. 
Previous work identified a critical transition to irreversible steady-states that are diffusive at long times, for $\phi_0 > \phi_c$ \cite{Corte08,Milz13}. 

Significant attention has been devoted to this model under isotropic initial conditions and driving \cite{Menon09,Keim13,Xu13,Milz13,Schrenk15,Tjhung16,Hexner17b}. 
Here we probe the transient dynamics as the particles are driven towards a hard boundary. 
Following the sedimentation protocol of Ref.~\cite{Corte09}, each cycle has an additional step where all particles move down a distance $v_s$. 
Particles stop settling at the bottom of the simulation box, and any kicks into that wall are specularly reflected. 
We use periodic boundary conditions in the horizontal direction. 
We study the behavior at low sedimentation speed, $v_s \ll 16\phi_c D W / (\pi d^2 N)$, where $D$ is the coefficient of diffusion for a non-sedimenting system measured at $\phi = 2\phi_c$ \cite{Corte09}. 
In this regime, particle transport due to sedimentation is much slower than from diffusion, when compared over the vertical lengthscale $\pi d^2 N / (4 \phi_c W)$ \cite{Wang18}, which is the height of a bed of particles of density $\phi_c$.

\begin{centering}
\begin{figure*}[tb]
\includegraphics[width=1.0\linewidth]{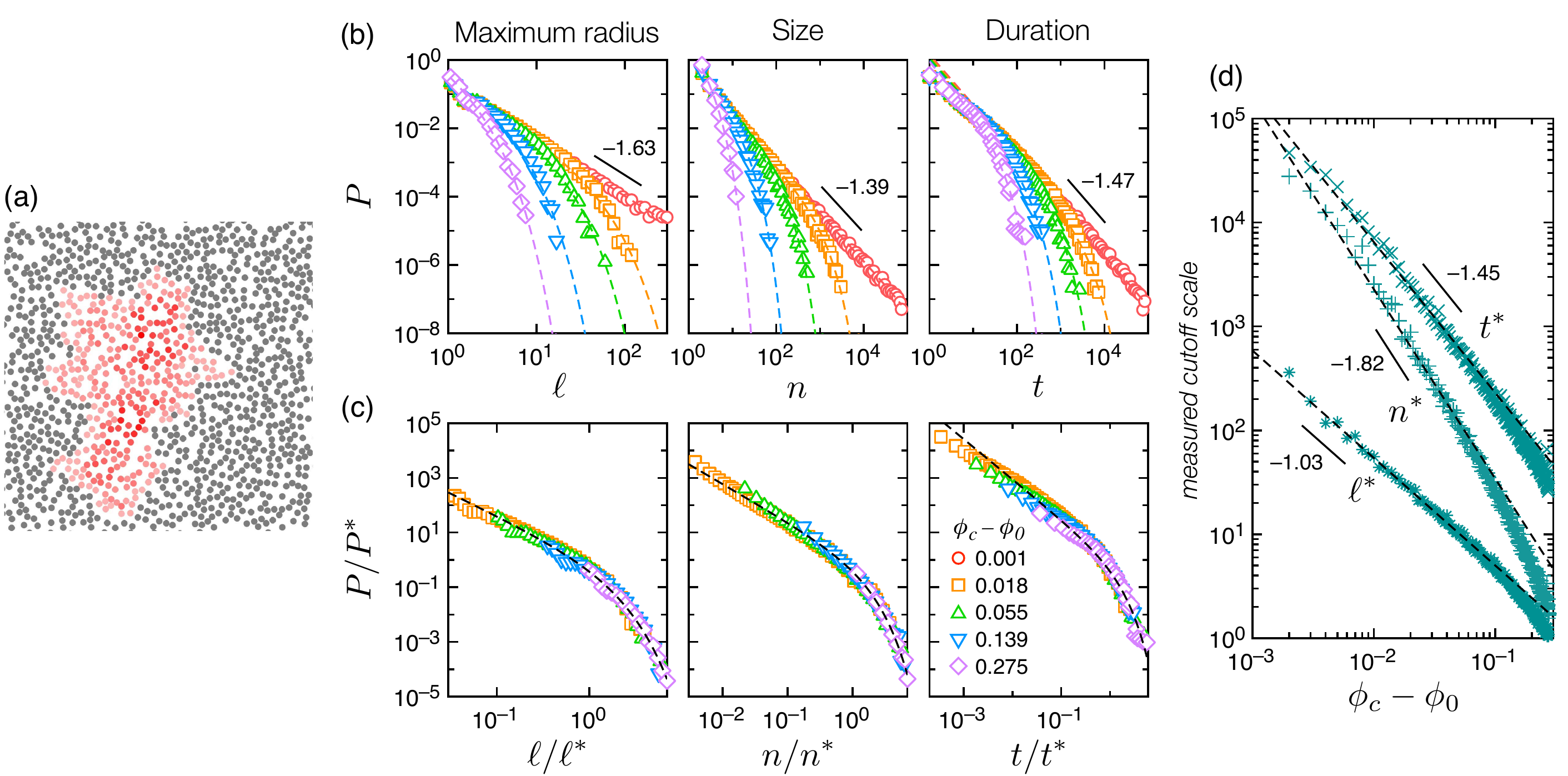}
\caption{
\textbf{Response to point perturbation.}
\textbf{(a)} Starting from a quiescent state, a perturbation may set off a chain reaction where many particles are activated before the system becomes quiescent again. 
Colored particles were active at some time during the avalanche, and the darker particles received more total kicks. 
\textbf{(b)} Histograms collected over many systems for the distance to the farthest activated particle, $\ell$, the number of activated particles, $n$, and the avalanche duration in cycles, $t$.
Solid lines: Fits to Eq.~(\ref{eq:hist}), where the measured exponent $\alpha$ is indicated in each panel. 
\textbf{(c)} The curves are approximately collapsed when scaled by the location of the exponential cutoff. 
\textbf{(d)} Value of the cutoff versus $\phi_c - \phi_0$. Each curve diverges as a power law, with an exponent that is distinct from $\alpha$ (see Table~\ref{tab:1}). 
All systems have $\epsilon = 0.5$, $W=H=400$, and $\phi_c = 0.375$. 
}
\label{fig:3}
\end{figure*}
\end{centering}

\textit{Compaction fronts---} Figure~\ref{fig:1}b shows a typical system evolution. 
As the particles settle at velocity $v_s$, a dense sediment builds up from the bottom wall, with its top surface propagating upwards at a velocity that we denote by $v_f$. 
If we assume that the upper region has constant density $\phi_0$ and the sediment has constant density $\phi_c$, then conservation of area dictates \cite{Brzinski18}: 
\begin{equation}\label{eq:v_f}
v_s \phi_0 = v_f (\phi_c - \phi_0) \ .
\end{equation}
To test this prediction, we first determine the value of $\phi_c$ corresponding to the particular $\epsilon$, $W$, and $H$ that was used in each simulation. 
We measure this in independent simulations without sedimentation, by gradually incrementing $\phi_0$ until we observe irreversible steady states. 
Figure~\ref{fig:1}c compares the observed front velocity scaled by the sedimentation velocity, $v_f/v_s$, versus the ratio $\phi_0/(\phi_c - \phi_0)$. 
The data are in good agreement with Eq.~\ref{eq:v_f}, supporting this straightforward picture for the front velocity. 


These considerations do not constrain the front profile. 
Figure~\ref{fig:2}a shows the horizontally-averaged particle density versus height at equal intervals in time, from a typical simulation. 
Shifting the curves onto one another, we find that the front shape is invariant in time (Fig.~\ref{fig:2}b). 
We measure the front width by fitting to a sigmoid: 
\begin{equation}\label{eq:phi_y}
\phi(y) = \phi_2 - \frac{\phi_2 - \phi_1}{1+e^{(y-y_f)/\Delta_f}} \ .
\end{equation} 
Although the observed plateaus at $\phi_1$ and $\phi_2$ are in general close to $\phi_0$ and $\phi_c$, we treat them as fitting parameters when measuring $\Delta_f$. 
Figure~\ref{fig:2}c shows that the measured front width depends strongly on $\phi_c - \phi_0$, where the $\phi_c$ are measured using independent simulations without sedimentation. 
We can also think of the top interface of the system in the steady state as a stationary front with $\phi_0=0$. 
We measure its width using Eq.~\ref{eq:phi_y} (with $\phi_1=0$) and we find the same trend as the transient measurements without any rescaling of parameters (Fig.~\ref{fig:2}c). 
Altogether, the data are consistent with a power law:
\begin{equation}\label{eq:Delta_vs_dphi}
\Delta_f \approx C (\phi_c - \phi_0)^{-\beta} \ . 
\end{equation}
with $\beta = 1.15 \pm 0.18$ and $C=0.24 \pm 0.06$. 

One may expect the kick size to affect the front width, since larger $\epsilon$ leads to a larger effective diffusion constant in the sediment. 
Surprisingly, we find the front width to be independent of $\epsilon$ in our simulations (Fig.~\ref{fig:2}d). 
Note that the system can discover denser reversible arrangements for smaller $\epsilon$. 
To account for this dependence of $\phi_c$ on $\epsilon$, we first measured $\phi_c$ independently in simulations without sedimentation, where we find that it varies from $0.20$ to $0.44$ as $\epsilon$ is decreased from $5$ down to $0.05$. 
We then set $\phi_0 = \phi_c(\epsilon) - 0.1$ for each of the simulations in Fig.~\ref{fig:2}d. 
This careful protocol reveals that $\Delta_f$ is independent of $\epsilon$ when $\phi_c - \phi_0$ is fixed. 

To look for any dependence on the system width $W$, we take the $\Delta_f$ measurements from Fig.~\ref{fig:2}c, divide them by the power-law fit, Eq.~\ref{eq:Delta_vs_dphi}, and plot this ratio in Fig.~\ref{fig:2}e. 
The data do not systematically increase with $W$, indicating that the interface is not rough \cite{Family90}.

\begin{table*}[tb]
\centering{}\begin{tabular}{l l l c c c }
\hline
\hline
&					& Expression			& DP   			& CDP/Manna		& Present work	\\ 
\hline
\textbf{Decay}
&Maximum radius		& $2\tau - 1$			& 1.536			& 1.560			& $1.63 \pm 0.10$	\\
&Size				& $\tau$ 				& 1.268			& 1.280			& $1.39 \pm 0.07$	\\
&Duration				& $\tau_t$				& 1.450			& 1.510			& $1.47 \pm 0.09$	\\
\hline
\textbf{Cutoff}	
&Maximum radius, $\ell^*$	& $1/(2\sigma)$ 		& 1.089			& 1.115			& $1.03 \pm 0.08$	\\
&Size, $n^*$				& $1/\sigma$			& 2.179			& 2.229			& $1.82 \pm 0.19$	\\
&Duration, $t^*$			& $1/\sigma_t$			& 1.297			& 1.225			& $1.45 \pm 0.14$	\\
\hline
\hline
\end{tabular}
\vspace{1em}
\caption{
\textbf{Comparison of critical exponents.} Values are shown for directed percolation (DP, obtained from Ref.~\cite{Munoz99}), conserved directed percolation (CDP/Manna, obtained from Ref.~\cite{Lubeck04}) and the present work using point perturbations in the isotropic random organization model. Greek notation matches that of Ref.~\cite{Munoz99}. 
}
\label{tab:1}
\end{table*}

\textit{Correlation lengthscale---}
It is natural to ask whether the finite interface thickness is a manifestation of a growing correlation lengthscale in the bulk. 
For random organization, 
Tjhung and Berthier \cite{Tjhung16} reported static and dynamic lengthscales with exponents of $0.73 \pm 0.04$ and $0.77 \pm 0.06$ respectively, and a hyperuniform lengthscale with exponents $0.76$ and $1.23$ when approaching $\phi_c$ from below and above, respectively \cite{Tjhung15}. 
Hexner and Levine reported a hyperuniform lengthscale with an exponent of $0.8$ for noiseless systems \cite{Hexner15} and $1.1 \pm 0.1$ when noise is present \cite{Hexner17b}. 
However, it is not a priori clear which of these exponents might be related to the diverging front width that we observe. 

One intuitive method to probe a diverging lengthscale is to perturb the system at a point and measure the characteristic radius of the affected region. 
We start by initializing random systems of density $\phi_0 < \phi_c$ in a square box with $W = H = 400$ and running the random organization model (with $v_s=0$) until they reach a reversible state.
Then, we give one particle a random kick. 
If it collides with another particle, we call this an ``avalanche'', and we evolve the system until it reaches another reversible state \cite{Perkovic95}. 
Figure \ref{fig:3}a shows an example, where the red particles were active at some time during the avalanche. 
For each avalanche, we measure: (i) the distance $\ell$ from the initial perturbation to the farthest final position of all affected particles, (ii) the size of the avalanche $n$, given by summing over all cycles the number of particles that are active in each cycle, and (iii) the duration $t$ of the avalanche in cycles. 

To build up statistics, we generate up to $100$ reversible states for each value of $\phi_0$; each is used in $10^3$ tests where we select one particle at random as the site of the perturbation. 
Histograms of $\ell$, $n$, and $t$ are shown in Fig.~\ref{fig:3}b for various $\phi_c - \phi_0$. 
We find good fits to the function:
\begin{equation}\label{eq:hist}
P(x) = A x^{-\alpha} \exp(-x/x^*) \ ,
\end{equation}
where $\alpha$ is determined by fitting a power law to the curve that is closest to the critical state, and $A$, $x^*$ are then fit for each curve. 
The data can be collapsed onto master curves by scaling the histograms by $x^*$ and $P^* = P(x^*)$ (Fig.~\ref{fig:3}c). 
We find good collapses for $\ell$ and $n$ but only an approximate collapse for $t$. 

To probe the variation of $\ell^*$, $n^*$, and $t^*$ with density, we generate additional histograms at over $200$ densities, 
and we measure the location of the exponential cutoff by fitting to Eq.~\ref{eq:hist}. 
The results are shown in Fig.~\ref{fig:3}d. 
The data diverge as $\phi_c$ is approached from below, with a different exponent for each quantity. 
Table~\ref{tab:1} summarizes the six measured exponents from Fig.~\ref{fig:3}b,d. 

The exponents from these avalanche should be shared with other models in the same universality class, which previous work has argued is either directed percolation (DP) or conserved directed percolation (CDP) \cite{Corte08,Menon09,Tjhung15,Tjhung16}. 
Our results are consistent with DP or CDP and cannot distinguish between the two. 
Going beyond these studies, here we propose that the largest radial extent of a `typical' avalanche, $\ell^*$, is governed by the exponent $1/(2\sigma)$ from DP or CDP. 
The numerical value is also close to the exponent for the lengthscale $\xi_2$ in Ref.~\cite{Hexner17b}, which shares the intuitive property of being the ``farthest'' distance of influence of re-organization events, and was measured to be $1.1 \pm 0.1$ in the Manna model \cite{Hexner17b}.

\textit{Connecting the correlation lengthscale $\ell^*$ to the interface thickness---} 
Returning to the original problem of propagating irreversibility fronts, we suggest that $\ell^*$ should be central to setting the interface thickness $\Delta_f$ (Fig.~\ref{fig:2}). 
In the low-sedimentation speed regime probed here, the interface is continually perturbed from below by particles in the active phase; these perturbations create avalanches that have the net effect of transporting particles upwards into the quiescent phase. 
The longest lengthscale of these disturbances should be set by $\ell^*$, which itself is set by the proximity of $\phi_0$ to the critical fraction, $\phi_c$.  
We test this intuitive picture by comparing the exponents in the two cases. 
We measure the exponent for $\ell^*$ to be $1.03 \pm 0.08$ (Fig.~\ref{fig:3}d), which is within the error bars of the exponent for the interface thickness, $1.15 \pm 0.18$ (Fig.~\ref{fig:2}c).

\textit{Discussion---}
Here we have observed an interface between reversible and irreversible phases in a model of a cyclically-sheared suspension, and we demonstrated the divergence of its thickness in the vicinity of a nonequilibrium critical point. 
Two properties of the interface place it in contrast with other non-equilibrium systems. 
First, it propagates with constant thickness (Fig.~\ref{fig:2}b), unlike many interfacial growth phenomena that are captured by Poisson-like growth or the Kardar-Parisi-Zhang universality class \cite{Family90,Yunker13}. 
Second, it is not observed to roughen (Fig.~\ref{fig:2}e), unlike what is observed in the two-dimensional Ising model \cite{Mon90}. 
Interestingly, the non-equilibrium phenomenology we observe has some similarities with an equilibrium fluid near a critical point: Both systems exhibit a diverging interface thickness that can be attributed to a diverging lengthscale in the bulk \cite{Cahn58,Huang69}. 
The observed density profile (Fig.~\ref{fig:2}b) is also consistent with the mean-field prediction in a van der Waals fluid \cite{Rowlinson13}. 
Nevertheless, the driving forces are clearly different --- diffusion only occurs for particles that overlap in our system, so that geometry plays a central role. 

Our results also share general features with dynamic jamming fronts, which arise in settings ranging from iceberg-choked fjords~\cite{Peters15} to water and cornstarch suspensions \cite{Roche13}. 
Such dynamic fronts develop when a collection of grains is impacted, creating a jammed region that grows as it amasses more grains on its boundary \cite{Waitukaitis12,Burton13}. 
Recent experiments measured a finite interfacial thickness between a dynamically jammed mass and its quiescent surroundings~\cite{Waitukaitis13}, and they showed that this thickness diverges as the dilute phase approaches the jamming density. 
They rationalized these findings by appealing to a diverging correlation length at the jamming point \cite{Ellenbroek06,Olsson07,Olson-Reichhardt10}. 
Here we observe a similar phenomenology in random organization under a slow external drive. 
This connection is perhaps surprising; in our system, particles in the front are continually activated into a diffusing state. 
One might expect this diffusion rate to influence the front width. 
Instead, we find the interfacial thickness is tied to geometric parameters through $\phi_c - \phi_0$, independent of dynamic parameters such as $\epsilon$ (Fig.~\ref{fig:2}c,d). 

This connection with dynamic jamming may prompt one to ask whether front formation could serve as an organizing principle among a broader set of nonequilibrium systems. 
The essential features underlying front formation appear to be: (\textit{i}) a critical transition between a dilute phase and a dense, incompressible phase
and (\textit{ii}) a process that compacts the system locally or at a boundary. 
These features might be found in active particle systems \cite{Henkes11}, which can form interfaces through motility-induced phase separation in which dense, fluid-like regions are surrounded by dilute, gas-like regions~\cite{Fily12,Siebert18,Partridge19}. 
Future work should investigate whether such interfaces share the phenomenology studied here. 
\begin{acknowledgments}
We thank Daniel Hexner for helpful discussions. 
J.W. and J.D.P. gratefully acknowledge the Donors of the American Chemical Society Petroleum Research Fund for partial support of this research. 
J.M.S. acknowledges NSF-DMR-CMMT-1832002 for financial support. 
This research was supported in part through computational resources provided by Syracuse University, including assistance from Larne Pekowsky under NSF award ACI-1541396. 
\end{acknowledgments}

%




\end{document}